\documentclass[twocolumn,trackchanges]{aastex631}

\usepackage{orcidlink}
\usepackage{graphicx}
\usepackage{bm}
\usepackage{hyperref}
\usepackage{dcolumn}
\usepackage{float}
\usepackage{xcolor}
\usepackage{multirow}
\usepackage{amsmath}

\newcommand{\Omk}{$\Omega_{k}$} %
\newcommand{\mnu}{$m_{\nu}$} 
\newcommand{\wO}{$w_{0}$} 
\newcommand{\wa}{$w_{\rm a}$} 
\newcommand{\As}{$A_{\rm s}$}
\newcommand{\gam}{$\gamma$}
\newcommand{\AIA}{A_{\rm IA}}
\newcommand{\Omegam}{\Omega_{\rm m}}
\newcommand{\baseplusmodel}{$w_{0}w_{\rm a}{\rm CDM}$}

\newcommand{\bfk}{\boldsymbol{k}}
\newcommand{\bfx}{\boldsymbol{x}}
\newcommand{\Cov}{{\rm Cov}}
\newcommand{\DA}{D_{\rm A}}

\newcommand{\wt}{\widetilde}
\newcommand{\Pgg}{P_{\rm gg}}
\newcommand{\PEE}{P_{\rm EE}}
\newcommand{\PgE}{P_{\rm gE}}
\newcommand{\wtPgg}{\wt{P}_{\rm gg}}
\newcommand{\wtPEE}{\wt{P}_{\rm EE}}

\newcommand{\be}{\begin{equation}}
\newcommand{\ee}{\end{equation}}

\begin{document}

\title{Squeezing Full-Shape Dynamical Dark Energy Constraints with Galaxy Alignments}

\author[0000-0001-7352-6175]{Junsup Shim}
\affiliation{Academia Sinica Institute of Astronomy and Astrophysics (ASIAA), No. 1, Section 4, Roosevelt Road, Taipei 106216, Taiwan}

\author[0000-0002-8942-9772]{Teppei Okumura}
\affiliation{Academia Sinica Institute of Astronomy and Astrophysics (ASIAA), No. 1, Section 4, Roosevelt Road, Taipei 106216, Taiwan}
\affiliation{Kavli IPMU (WPI), UTIAS, The University of Tokyo, Kashiwa, Chiba 277-8583, Japan}
\author[0000-0002-4016-1955]{Atsushi Taruya}
\affiliation{Yukawa Institute for Theoretical Physics, Kyoto University, Kyoto 606-8502, Japan}
\affiliation{Kavli IPMU (WPI), UTIAS, The University of Tokyo, Kashiwa, Chiba 277-8583, Japan}



\begin{abstract}

Recent $2-4\sigma$ deviations from the Cosmological Constant $\Lambda$ suggest that dark energy (DE) may be dynamical, based on baryon acoustic oscillations and full-shape galaxy clustering (FS GC) analyses. This calls for even tighter DE constraints to narrow down its true nature.
In this Letter, we explore how galaxy intrinsic alignments (IA) can enhance the FS GC-based DE constraints, using Fisher forecasts on various extensions of dynamical DE models, including scenarios with curvature, massive neutrinos, and modified gravity. Incorporating IA improves the DE Figure-of-Merit by $42-57\%$ and tightens the primordial power spectrum amplitude constraints by $17-19\%$. Our findings highlight IA's potential as a valuable cosmological probe complementary to GC.

\end{abstract}

\keywords{Dark Energy (351) --- Cosmological parameters from LSS (340) --- Astrostatistics strategies (1882)}


\section{Introduction} \label{sec:intro}

Recent cosmological analyses by the Dark Energy Spectroscopic Instrument (DESI) \citep{desi+16} suggested a statistical preference for dynamical dark energy (DE) over the static Cosmological Constant $\Lambda$ \citep{desiBAO+24, desiFS+24, desiFSMG+24}, broadening opportunities for evolving DE models \citep[see][ and references therein]{copeland+06,tsujikawa13,amendola&tsujikawa15} to explain the observed accelerated expansion of the Universe \citep{riess+98,perlmutter+99}.
However, ultimately identifying the true nature of DE further requires more accurate and precise constraints, which necessitates the maximal extraction of cosmological information from observational data.

In the spirit of tightening cosmological constraints, the major source of cosmological information has been characteristic clustering features imprinted in the large-scale galaxy distribution, e.g., the baryon acoustic oscillation (BAO) \citep{peebles&yu70,eisenstein&hu98} and redshift-space distortion (RSD) \citep{kaiser87, hamilton92}. They serve as effective tools for measuring the expansion and growth rates of the Universe, providing constraints on DE and modified gravity (MG) models \citep{peacock01,seo&eisenstein03,tegmark+04,okumura+08,guzzo+08,beutler+12,blake+12,reid+12,samushia+13,beutler+14,aubourg+15,okumura+16,alam+17,beutler+17,gilmarin+17,hawken+17,hou+21,aubert+22}.

While such clustering analyses focus on the spatial correlations of galaxy positions, the alignment of galaxy shapes has emerged as another valuable cosmological observable \citep[see][for reviews]{kirk+15, troxel&mustapha15}.
Apart from the apparent galaxy shape alignments due to the weak gravitational lensing, or cosmic shear \citep{blanford+91,kaiser92, miralda91,bartelmann&schneider01}, the intrinsic alignments (IA) of galaxy shapes \citep{brown+02, mandelbaum+06, hirata+07, okumura+09, tonegawa+22, tsaprazi+22, OT23, zhou+23} also carry cosmological information since the spatial correlations of IA follow the tidal field induced by the underlying matter distribution.
Thus, IA ultimately probes the same matter density field as galaxy clustering. However, unlike galaxy clustering, IA disentangles the growth and expansion rate measurements, leveraging its insensitivity to redshift space distortion \citep{okumura+24}. Consequently, IA can tighten constraints on the expansion and growth rates when combined with galaxy clustering \citep{TO20,OT22,OT23,xu+23}. In addition, IA can potentially offer clues to the early universe physics, including primordial gravitational wave, non-gaussianity, and magnetic fields \citep{schmidt&jeong12,chisari&dvorkin13,akitsu+21,biagetti&orland020,kurita&takada23,philcox+24,saga+24}.

In this Letter, we investigate how effectively the IA of galaxies can improve the DE constraints obtained from galaxy clustering (GC). More specifically, we perform a Fisher forecast to assess the gain within the full-shape (FS) framework, where the broadband shape of the matter power spectrum is further exploited. This approach differentiates our work from other cosmological inferences leveraging IA and marks the first attempt to investigate the benefit of the FS information of IA in DE constraints. We also forecast the primordial power spectrum amplitude directly encoded in the FS GC and IA signal \citep[see][ for other parameter constraints]{shim+25}.
We explore a broader range of dynamical DE models -- including extensions involving curvature, massive neutrinos, and MG -- beyond those investigated in the recent DESI's analyses \citep{desiBAO+24, desiFS+24,desiFSMG+24}.
While such model extensions relax DE constraints \citep{choudhury&okumura24}, we demonstrate that IA significantly tightens DE constraints across all models, compared to constraints obtained from FS GC alone.

\section{IA and GC statistics} \label{sec:model}

We extract cosmological information from GC and IA by utilizing two-point statistics of galaxy density and shape fields.
The statistics for GC and IA are constructed from the fluctuation of number density, $\delta_{\rm g}(\bfx)$, and two-component ellipticity, $(\gamma_{+},\gamma_{\times})$.
In Fourier space, the former is related to the underlying matter density field following \citep{kaiser87},
\be
    \delta_{\rm g}(\bfk,z) = (b_{\rm g}(z)+f(k,z)\mu^{2})\delta_{\rm m}(\bfk,z)\ ,
    \label{eq:delta}
\ee
where $b_{\rm g}$ expresses the linear bias relation between galaxies and matter density field \citep{kaiser84} and $f\mu^{2}$ accounts for the anisotropies along the line-of-sight due to galaxy's peculiar motion \citep{kaiser87}, with $f$ and $\mu$ denoting the linear growth rate and directional cosine between the wavevector and line-of-sight. The presence of massive neutrinos introduces the scale-dependence to the linear growth rate \citep{kiakotou+08, boyle&komatsu+18} due to their free-streaming motions \citep{lesgourgues&pastor06}.

The orientations of galaxy shapes projected onto the sky can be characterized by ellipticities, defined with the minor-to-major axis ratio $q$ as,
\be
\gamma_{(+,\times)}(\bfx,z) = \frac{1-q^2}{1+q^2} \left( \cos{2\theta},\sin{2\theta} \right),
\ee
with $\theta$ measuring the angle between the galaxy's major and reference axes. In Fourier space, the ellipticity fields then can be re-expressed using rotation-invariant E/B components \citep{stebbins+96,kamionkowski+98,crittenden+02} via $\gamma_{\rm E}(\bfk,z) + i\gamma_{\rm B}(\bfk,z) = e^{-2i\phi_k}\left\{\gamma_{+}(\bfk,z)+i\gamma_{\times}(\bfk,z) \right\}$ with $\phi_{k}={\rm arctan}(k_y/k_x)$.

Adopting the linear alignment model \citep{catelan+01,hirata&seljak04}, where IA arises as a linear response to the gravitational tidal field generated by the surrounding large-scale structures, the ellipticity fields are given,
\be
\gamma_{(+,\times)}(\bfk,z) = b_K(z) \left( k_x^2-k_y^2, 2k_xk_y \right)\frac{\delta_{\rm m}(\bfk,z)}{k^2}.
\label{eq:gamma_+x}
\ee
Here, we used the Poisson equation to replace gravitational potential with the matter density fluctuation. The shape bias, $b_{K}$, quantifies the response of galaxy shapes to the tidal field, further described by introducing the amplitude of IA \citep{joachimi+11, shi+21a, shi+21b, kurita+21, inoue+24}, $\AIA$, as $b_K(z) = - 0.01344 \AIA{}\Omegam/D(z)$, with $D(z)$ being the linear growth factor. Then, Eqn.~\ref{eq:gamma_+x} yields $\gamma_{E}$ to become the only non-vanishing ellipticity component,
\be
\gamma_{\rm E}(\bfk,z)=b_K(z) (1-\mu^2) \delta_{\rm m}(\bfk,z)\ ,
\label{eq:gamma_E}
\ee
where anisotropies along line-of-sight with $\mu$ appear, similarly in Eqn.\ref{eq:delta}, because galaxy shapes are projected on the celestial plane perpendicular to the line-of-sight.

From these observables, we compute three power spectra, i.e., auto-power spectra of $\delta_{\rm g}$ and $\gamma_{\rm E}$, and their cross-power spectrum, respectively expressed as
\be
    P_{\rm gg}(k,\mu,z) =(b_{\rm g}+f(k,z)\mu^{2})^{2}P_{\rm m}(k,z),\label{eq:pgg} 
\ee
\be
    P_{\rm gE}(k,\mu,z) = b_K(1-\mu^{2})(b_{\rm g}+f(k,z)\mu^{2})P_{\rm m}(k,z),\label{eq:pgE} 
\ee
\be
    P_{\rm EE}(k,\mu,z) = b_K^{2}(1-\mu^{2})^{2}P_{\rm m}(k,z),\label{eq:pEE} 
\ee
where $P_{\rm m}$ is the linear matter power spectrum \citep[see][for their configuration-space counterparts]{OT20b}. Accounting for the Alcock-Paczynski effect \citep{alcock&paczynski79}, induced by the mismatch between the fiducial and true cosmologies, observed power spectra become,
\be
P_{i}^{\rm obs}\left(k_\perp^{\rm fid},k_\parallel^{\rm fid},z \right) = 
\frac{H(z)}{H^{\rm fid}(z)}\left\{ \frac{D_{\rm A}^{\rm fid}(z)}{\DA(z)} \right\}^2
P_{i}\left(k_\perp,k_\parallel, z \right), 
\label{eq:Pi_AP}
\ee
where $(k_{\perp}^{\rm fid}, k_{\parallel}^{\rm fid})=(\DA(z)/D_{\rm A}^{\rm fid}(z)k_{\perp}, H^{\rm fid}(z)/H (z)k_{\parallel})$ and $i=\rm (gg, gE, EE)$. Here, the wavenumber $k$ is expressed with $(k_\perp,k_\parallel) = k(\sqrt{1-\mu^2},\mu)$. The quantities with a superscript `fid' are those in an assumed cosmological model. 
In FS analyses, cosmological parameter constraints are extracted not only from the geometric/dynamical quantities, such as $H(z), D_{\rm A}(z)$, and $f(k,z)$, but also from the shape of the matter power spectrum, which provides additional cosmological information.

\section{Forecasting cosmological constraints} \label{sec:fisher}

Using the three observed power spectra as cosmological probes, we perform Fisher analysis to quantify improvement with IA in constraining parameters, $\boldsymbol{\theta}=({\rm ln} (10^{10}A_{\rm s}), w_{0}, w_{\rm a})$, where \As{} stands for the primordial amplitude of the power spectrum and \wO{} and \wa{} are the DE equation of state (EOS) in CPL-parametrization \citep{chevallier&polarski01,linder03}. 
For $P_{i}=(\Pgg,\PgE,\PEE)$, the Fisher matrix can be evaluated via,
\begin{align}
F_{\alpha\beta}(z) = & \frac{V_s}{4\pi^2} \int^{k_{\rm max}}_{k_{\rm min}} dk k^2 \int ^{1}_{-1}d\mu \nonumber \\ & 
\times \sum_{i,j} \frac{\partial P_i(k,\mu,z)}{\partial\theta_\alpha}
\left[ \Cov^{-1}\right]_{ij}\frac{\partial P_j(k,\mu,z)}{\partial\theta_\beta},
\label{eq:Fisher_matrix}
\end{align}
where $V_{s}$, $k_{\rm min}$, and $k_{\rm max}$ represent the survey volume, minimum and maximum wavenumbers for the analysis.
Note that our Fisher matrix also includes constraints on other free parameters, including five fiducial parameters, as well as curvature (\Omk), massive neutrino (\mnu), and modified gravity (\gam) parameters. For details on these parameter constraints, we refer readers to \citet{shim+25}.
The Gaussian covariance matrix of the observed power spectra is defined ${\rm Cov}_{ij}(k,\mu,z)=\langle P_{i}P_{j}\rangle-\langle P_{i}\rangle\langle P_{j}\rangle$. When jointly utilizing IA and GC, it becomes 
\be
\Cov_{ij} =
\left[
\begin{array}{ccc}
2 \{ \wtPgg\}^2 & 2\wtPgg\PgE & 2 \{ \PgE\}^2 \\
2\wtPgg\PgE & \wtPgg \wtPEE + \{ \PgE\}^2 & 2\PgE\wtPEE \\
2 \{ \PgE\}^2 & 2\PgE\wtPEE & 2 \{ \wtPEE\}^2 \\
\end{array}
\right]\ ,
\label{eq:covariance}
\ee
with $\wt{P}_{\rm gg}$ and $\wt{P}_{\rm EE}$ denoting auto-power spectra with the Poisson shot noise, $\wtPgg = \Pgg + 1/n_{\rm g}$ and 
$\wtPEE = \PEE + \sigma_\gamma^2/n_{\rm g}$, respectively.
Here, $n_{\rm g}$ is the mean galaxy number density, and $\sigma_{\gamma}$ is the shape-noise due to the scatter in the intrinsic shapes and measurement uncertainties.

\section{Setup and results} \label{sec:results}

For our forecast, we consider a deep galaxy survey like Prime Focus Spectrograph (PFS) \citep{pfs+14}, which scans a wide range of redshift $0.6\le z<2.4$, targetting to probe the redshift evolution of DE. Furthermore, it allows precise IA measurements, aided by the Hyper Supreme-Cam imaging survey \citep{hsc+18,hsc2+18}. We briefly discuss forecast results assuming a Euclid-like wide survey \citep{euclid+11} to comment on how our results change depending on survey geometries, e.g., deep PFS-like and wide Euclid-like surveys.
Parameters characterizing the PFS-like and Euclid-like surveys and their observed galaxies (i.e., $V_{\rm s}, \bar{n}_{\rm g}$, and $b_{\rm g}$) are set following Tables 2 in \citet{pfs+14} and \citet{blanchard+20}, respectively. We set $A_{\rm IA}=18$ and $\sigma_{\gamma}=0.2$, assuming the IA estimator developed for emission-line galaxies (ELGs) \citep{shi+21a} and high-quality shape information \citep{hsc+18,hsc2+18}, respectively. Note that this estimator effectively measures shapes similar to the host dark matter halo shapes from the light distribution around ELGs within aperture radius. The maximum wavenumber for the analysis is $k_{\rm max}=0.2h{\rm Mpc}^{-1}$, with $h$ defined as the present-day Hubble constant divided by $100{\rm km}/{\rm s}/{\rm Mpc}$. The Planck-15 compressed likelihood \citep{planck+15} is included as a CMB prior.

\begin{figure}
\includegraphics[width=\columnwidth]{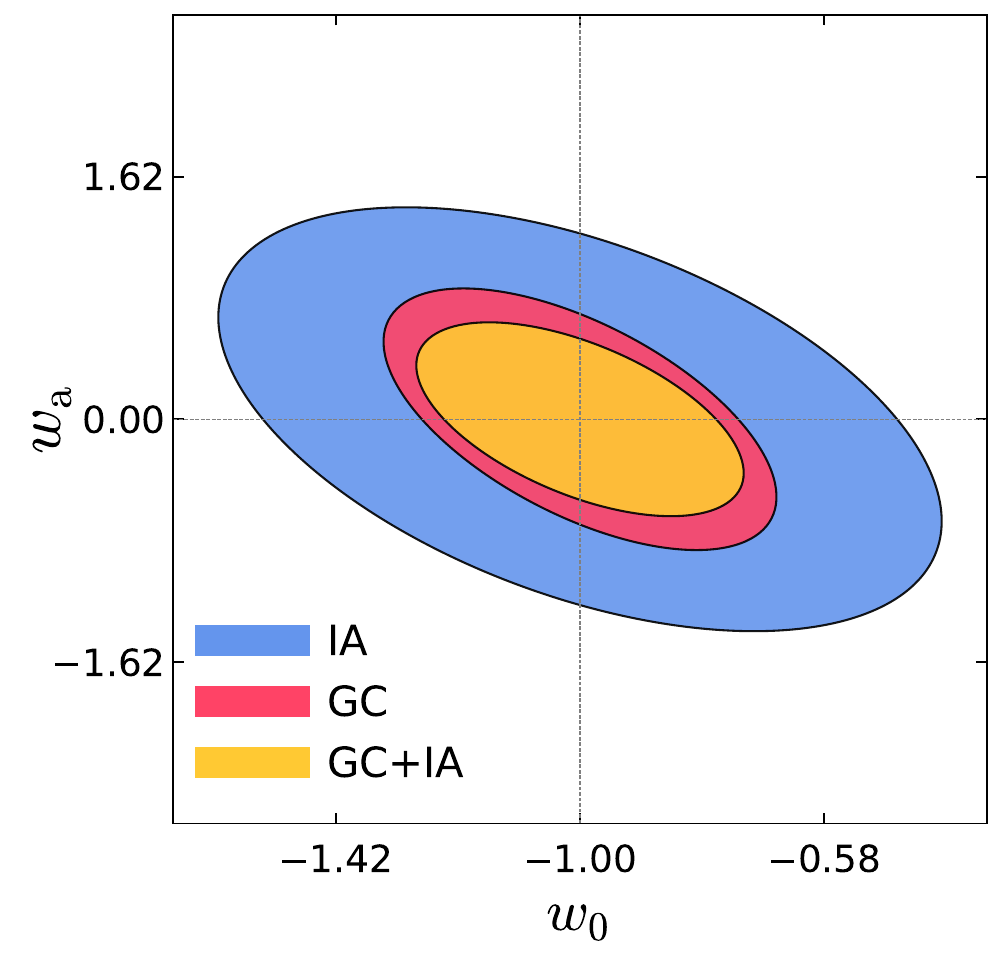}
\caption{2D-confidence ellipses for DE EOS parameters of the most extended dynamical DE model, \baseplusmodel{}+(\Omk, \mnu, \gam) assuming a PFS-like survey. Contours represent 1-$\sigma$ confidence regions and CMB prior is included. Contour for `IA' is obtained from $P_{\rm EE}$ and $P_{\rm gE}$.}\label{fig:1}
\end{figure}
In Fig.~\ref{fig:1}, we show the 2D constraints on the DE parameters obtained from IA, GC, and their combination for the most extended dynamical DE model, i.e. \baseplusmodel{}+(\Omk, \mnu, \gam). It demonstrates that IA can provide cosmological constraints on the DE parameters, although its constraining power is relatively weaker than that of GC. However, when IA is combined with GC, the ellipse for GC shrinks noticeably, implying an improvement in the DE constraints due to extra information from IA. In such a case, the Figure-of-Merit (FoM) for the DE parameters increases roughly by $57\%$, where it is defined as ${\rm FoM} \equiv \sqrt{{\rm det}({\tilde{\boldsymbol{ F}}_{w_{0}w_{\rm a}}})}$ with a sub-Fisher matrix, $\tilde{{\boldsymbol{F}}}_{w_{0}w_{\rm a}}$. We obtain the sub-Fisher matrix by marginalizing the full-Fisher matrix over parameters other than \wO{} and \wa{} \citep{albrecht+06}. Moreover, adding IA also impacts the direction of degeneracy between the DE parameters. This is because IA exhibits a different degeneracy direction from that of GC, suggesting the potential of IA to weaken the parameter degeneracy.
\begin{figure}
\includegraphics[width=\columnwidth]{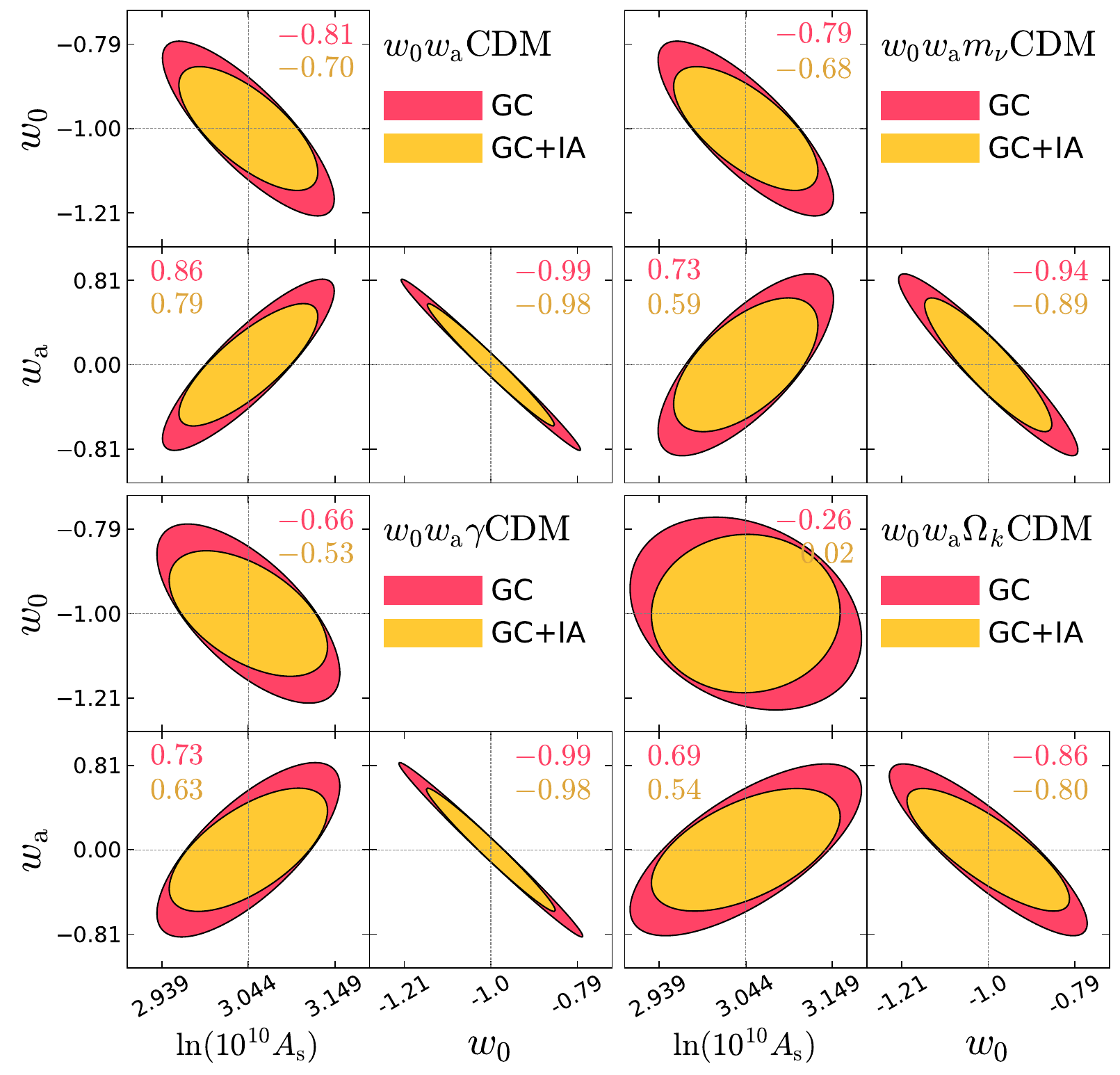}
\caption{2D-confidence ellipse contours for the DE EOS and \As{} parameters for \baseplusmodel{} and its one-parameter extension, including CMB prior. 1$\sigma$-confidence contours are shown. 
Correlation coefficient, $\rho_{XY}$ for a pair of cosmological parameters, $X$ and $Y$, is calculated as $\rho_{XY}=C_{XY}/\sqrt{C_{XX}C_{YY}}$ using the error covariance matrix, $\boldsymbol{C}_{XY}=(\tilde{\boldsymbol{F}}_{XY})^{-1}$.
}\label{fig:2}
\end{figure}

Such improvement with FS IA in DE constraints is not limited to the most extended dynamical DE model. Fig.~\ref{fig:2} exemplifies the cosmological benefit of IA in constraining the simplest \baseplusmodel{} model and its one-parameter extensions. We also display 2D constraints involving \As.
For the one-parameter extended models, ellipses become larger than those in the simplest model. This is because adding an extra cosmological parameter newly introduces additional parameter degeneracies, leading to a larger uncertainty.
However, in all four models, IA significantly tightens 2D constraints involving the DE and \As{} parameters. Relative to the GC-only cases, the gains in DE FoM with IA range $44-47\%$ in these models. Similarly, the contours involving \As{} also shrink substantially with IA, suggesting the efficacy of IA in constraining \As{}.
It is worth noting that IA significantly decreases correlation coefficients for \wO-\As{} and \wa-\As{}, likely indicating the reduced degeneracies for those parameter pairs. For the 2D constraints and degeneracies of other cosmological parameter pairs, we direct the readers to Fig.7 in \citet{shim+25}.

\begin{figure}
\includegraphics[width=\columnwidth]{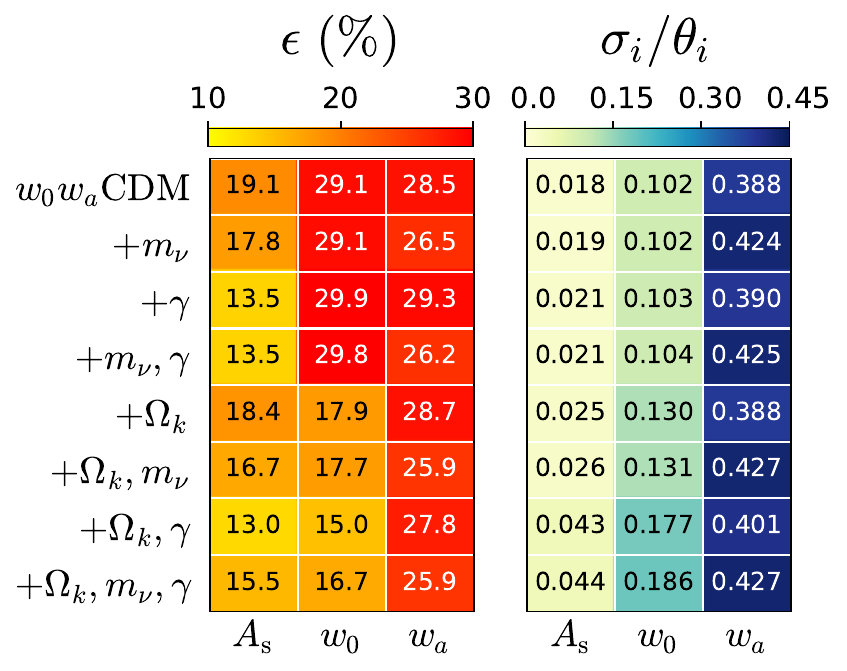}
\caption{Improvement in 1D-marginalized constraints with IA (left) and FS joint constraints (right), assuming a PFS-like survey with CMB prior included. Joint constraints as fractional errors, $\sigma_{i}/\theta_{i}$, are shown. For \wa{}, we show their actual errors, $\sigma_{i}$, since their fiducial values, $\theta_{i}$, are zero.}\label{fig:3}
\end{figure}

Let us now examine the improvement in 1D-marginalized constraints with IA. As can be seen in the left panel of Fig.~\ref{fig:3}, for all dynamical DE models investigated, DE and \As{} constraints are substantially improved with IA. The improvement is particularly significant for \wa{}, tightened at least by $25\%$ in all cases. The \wO{} constraints are improved nearly by $30\%$ in flat dynamical DE models, but noticeably less in nonflat models. Overall, the DE FoM improvement ranges from $42\%$ to $57\%$, depending on extra parameters added to the \baseplusmodel{} model.
Such improvements in FS DE constraint are noteworthy, as they are comparable to those predicted in geometric/dynamical analysis \citep{TO20, OT22}, where the absence of FS information results in inherently weaker constraints, leaving significantly more room for IA to improve cosmological constraints.

The reduction in marginalized error for \As{} is $17-19\%$-level in DE models without MG, but it becomes $1-7\%$ weaker in \gam{}-added models.
Such model-dependent improvements can be attributed to the increased parameter degeneracies in the joint analysis, as shown by the changes in correlation coefficients in Tab.~\ref{tab:1}. Adding IA information noticeably increases the correlation coefficients for \wO-\Omk, \wa-\mnu, and \As-\gam, implying that IA strengthens their degeneracies. Thus, the contribution of IA becomes less effective in the presence of a particular extra parameter, e.g., \Omk, \mnu, and \gam, due to its strengthened degeneracies with the DE EOS or \As{} parameters.

In the right panel of Fig.~\ref{fig:3}, we show the marginalized errors in the joint analysis. The marginalized error on \As{} is percent-level in all models. At the same time, \wO{} and \wa{} can be constrained at $10-19\%$ and $39-43\%$ levels, respectively. We note that our joint FS DE constraints are about $2-4$ times tighter than those from the joint geometric/dynamical analysis with a similar survey setup \citep{OT22}.
As expected, the fractional errors increase when more extra cosmological parameters are added to the \baseplusmodel{} model. The parameter degeneracy-dependent trend is again found in the marginalized errors.
For instance, the marginalized errors on \wO{} become noticeably larger in nonflat models, at least by $30\%$ compared to those in flat cases. This can be explained by \wO{}'s strongest degeneracy with \Omk{} than \mnu{} or \gam, which remains the strongest even after combined with IA. Thus, the \wO{} constraints become substantially worse in the presence of curvature.
Similarly, the existence of massive neutrinos worsens \wa{} constraints roughly by $10\%$ compared to the cases without the massive neutrinos. Again, this likely originates from the tightest correlation of \wa{} with \mnu{}. A similar explanation can be applied to the \As{} constraints becoming weaker in models with \gam{}.

\begin{table}
\centering
\caption{Correlation coefficients between cosmological parameters to be constrained (first row) and extra parameters (first column) when utilizing GC-only information and jointly with IA.}
\begin{tabular}{|c|c|c|c|c|c|c|}
\hline
& \multicolumn{2}{c|}{\As} & \multicolumn{2}{c|}{\wO} & \multicolumn{2}{c|}{\wa} \\
\cline{2-7}
& GC & +IA & GC & +IA & GC & +IA \\
\hline
\Omk & 0.67 & 0.68 & 0.42 & 0.62 & 0.08 & -0.06 \\
\mnu & 0.20 & 0.27 & 0.02 & -0.01 & -0.34 & -0.40 \\
\gam & -0.34 & -0.48 & -0.22 & -0.16 & 0.19 & 0.12 \\ 
\hline
\end{tabular}\label{tab:1}
\end{table}

\section{Conclusions}

Unveiling the true nature of DE has entered into a new era as DE may be dynamic rather than static, as suggested by the recent BAO \citep{desiBAO+24} and FS analyses \citep{desiFS+24, desiFSMG+24} by the DESI collaboration. Thus, tightening the current DE constraints will play a pivotal role in ultimately identifying DE.

In this Letter, we investigated the efficacy of IA in improving DE EOS and \As{} constraints, leveraging its cosmological information complementary to GC \citep{TO20, OT22, shim+25}.
For the first time, we have demonstrated that FS IA information significantly tightens the FS GC constraints on various extensions of the dynamical DE model.
Specifically, 1D-marginalized constraints on \wO{} and \wa{} are improved by $15-30\%$ and $26-29\%$, achieving $42-57\%$ of DE FoM gains.
Such enhancement with FS IA, relative to FS GC, represents a remarkable advancement
given that FS GC information can already substantially improve geometric DE constraints, e.g., DESI's FS GC \citep{desiFS+24} tightening DE FoM roughly by $20\%$ from its BAO-based constraints \citep{desiBAO+24}.

Our joint FS analysis forecasts $10\%$-level of marginalized (fractional) error on \wO{} in flat \baseplusmodel{} models. The constraints exacerbate in nonflat models, yielding $19\%$-level precision in the most extended dynamical DE model. The \wa{}-constraints are weaker, on average around $40\%$-level precision, becoming more uncertain in the presence of massive neutrinos.
For the simplest \baseplusmodel{} or \baseplusmodel{}+\mnu{} models, our joint constraints on \wO{} and \wa{} are $12-35\%$ and $5-40\%$ weaker than those from the DESI's FS GC results \citep{desiFS+24}, depending on their choices of supernovae type-Ia (SNIa) data. However, it should be noted that we only added FS IA to FS GC, whereas in \citet{desiFS+24} FS GC is combined with the BAO and SNIa results that more directly probe the nature of DE.
It is also worth noting that the joint analysis with IA makes DE constraints robust against adding MG to the flat dynamical DE mode. In such a model, FS GC-alone would lead to weaker DE constraints without further supplementary probes, e.g., weak gravitational lensing \citep{desiFSMG+24}.

We tested the robustness of our forecasts by examining how survey geometry and fiducial assumptions -- specifically the shape-noise level, IA amplitude, and maximum wavenumber -- affect the cosmological utility of IA.
For a Euclid-like wide survey with $\sigma_{\gamma}=0.3$ and $A_{\rm IA}=18$, the DE FoM improvement with IA is less pronounced than a deep PFS-like case, e.g. $21-24\%$ for flat dynamical DE models. However, their 1D-marginalized DE constraints are tighter than those for PFS-like cases approximately by $47\%$. Such tighter constraints are available due to the larger survey area in a wide Euclid-like survey, whereas its milder DE FoM improvement is attributed to the larger $\sigma_{\gamma}$ for a Euclid-like case.

Improvements in DE constraints are found to depend monotonically on both the shape noise level and IA amplitude \citep{TO20, OT22}. The DE FoM increases toward smaller $\sigma_{\gamma}$ and larger $A_{\rm IA}$. For example, for a PFS-like survey, reducing the shape noise from $\sigma_{\gamma}=0.20$ to 0.15 doubles the DE FoM, whereas the DE FoM is halved when increasing the shape noise to $\sigma_{\gamma}=0.30$. On the other hand, the DE FoM improvement rises to $74-98\%$ when adopting $A_{\rm IA}=24$, whereas it drops to $21-26\%$ for $A_{\rm IA}=12$.
Varying the maximum wavenumber $k_{\rm max}$ reveals a more nontrivial trend: the DE FoM generally increases as $k_{\rm max}$ decreases but exhibits oscillatory behavior, similar to the findings in \citep{taruya+11, shim+25}. For example, when limiting the analysis to $k_{\rm max}=0.1 h^{-1}{\rm Mpc}$, the DE FoM improvement with IA increases significantly to $75-82\%$. This reflects the relatively greater impact of IA in more conservative choices of $k_{\rm max}$, emphasizing the importance of IA within the linear theory description. Given the efforts in developing a nonlinear IA modeling based on perturbation theory, effective field theory, and simulations \citep{blazek+19,vlah+20,vlah+21,bakx+23,chen+24,maion+24,taruya+24}, it would also be interesting to investigate the benefit of FS IA by pushing to a larger $k_{\rm max}$ with such nonlinear IA modeling.

\section*{ACKNOWLEDGMENTS}
We thank an anonymous referee for helpful comments that helped improve the original manuscript. JS acknowledges the support by Academia Sinica Institute of Astronomy and Astrophysics. TO acknowledges the support of the Taiwan National Science and Technology Council under Grants No. NSTC 112-2112-M-001-034- and NSTC 113-2112-M-001-011-, and the Academia Sinica Investigator Project Grant (AS-IV-114-M03) for the period of 2025-2029. This work was supported in part by MEXT/JSPS KAKENHI Grants No. JP20H05861, No. JP21H01081 (AT). 

%


\bibliography{sample631}{}
\bibliographystyle{apj}



\end{document}